\documentclass{article}
\usepackage[latin9]{inputenc}
\usepackage[a4paper]{geometry}
\usepackage{amsmath}
\usepackage{amssymb}
\usepackage{esint}

\makeatletter
\newcommand{\lyxaddress}[1]{
\par {\raggedright #1
\vspace{1.4em}
\noindent\par}
}

\makeatother

\begin{document}

\title{Some aspects of affine motion and nonholonomic constraints. Two ways
to describe homogeneously deformable bodies}

\author{Barbara Go\l{}ubowska}

\maketitle

\lyxaddress{Institute of Fundamental Technological Research, Polish Academy of
Sciences $5^{B}$ Pawi\'{n}skiego str., 02-106 Warsaw, Poland e-mail:
bgolub@ippt.pan.pl}
\begin{abstract}
This paper has been inspired by ideas presented by V. V. Kozlov in
his works \cite{Koz_83,Koz_88}. In this paper our goal is to carry
out a thorough analysis of some geometric problems of the dynamics
of affinely-rigid bodies. We present two ways to describe this case:
the classical dynamical d'Alembert and variational, i.e., vakonomic
one. So far, we can see that they give quite different results. The
vakonomic model from the mathematical point of view seems to be more
elegant. The similar problems were examined by Jó\'{z}wikowski and
W. Respondek in their paper \cite{Joz_Res_13}. 
\end{abstract}

\section{General formulation}

The origin of the problem goes to the turn of XIX and XX centuries,
or even more precisely to the second half of XIX century. Let us begin
with reminding some elementary concepts. The configuration space of
mechanical systems will be a differential manifold $Q$ of dimension
$n$; local coordinates in $Q$, i.e., generalized coordinates will
be, as usual, denoted by $q^{1},\ldots,q^{n}$, or briefly by $q^{i}$.
The manifold of positions and velocities, i.e., the tangent bundle
over $Q$ will be denoted by $TQ$. It is a set-theoretical union
of all tangent spaces $T_{q}Q$, i.e., 
\begin{equation}
TQ=\underset{q\in Q}{\bigcup}T_{q}Q.\label{EQ1}
\end{equation}
Obviously, $T_{q}Q$ is the linear space of all apriori possible velocities
at $q$. Its dimension is $n$ as well, therefore $\dim TQ=2n$. For
any velocity vector at $q$, $v\in T_{q}Q$, we introduce its components
$v^{i}$ with respect to generalized coordinates $q^{i}$. This induces
coordinates $\left(q^{i},v^{i}\right)$ on $TQ$. The natural projection
of $TQ$ onto $Q$ will be denoted as usual: 
\begin{equation}
\tau_{Q}:TQ\rightarrow Q,\qquad\tau_{Q}\left(T_{q}Q\right)=\left\{ q\right\} .\label{EQ2}
\end{equation}
Strictly speaking, coordinates $q^{i}$ in $\left(q^{i},v^{i}\right)$
should be rather denoted by $q^{i}\circ\tau_{Q}$, however for the
obvious reasons of simplicity we use the traditional symbols $q^{i}$.
Geometry of the tangent bundle is interesting in itself and may be
considered as a mathematical background for the theory of the systems
of second-order differential equations \cite{AbrMar78,JJS_87,JJS91}.
Its advantage is to use only operational, directly measurable concepts
like coordinates and velocities. Any curve $\rho:\mathbb{R}\rightarrow Q$
may be canonically lifted to the curve $\rho^{\prime}:\mathbb{R}\rightarrow TQ$
without introducing any additional geometry to $Q$; just $\rho^{\prime}(t)$
is the tangent vector to $\rho$ at $\rho(t)$, a uniquely defined
element of the tangent space $T_{\rho(t)}Q$. And conversely, a curve
$\sigma:\mathbb{R}\rightarrow TQ$ is said to be integrable if there
exists a curve $\rho:\mathbb{R}\rightarrow Q$ such that $\sigma=\rho^{\prime}$.
At every point $v\in TQ$ there is a manifold of integrable vectors
$K_{v}\subset T_{v}TQ$ which $\tau_{Q}$ projects onto $v$, 
\begin{equation}
K_{v}=\left\{ X\in T_{v}TQ:T\tau_{Q}(X)=v\right\} .\label{EQ3}
\end{equation}
Only integrable vectors describe accelerations of motion; this description
is independent of anything like an affine connection or metric tensor
in $Q$. A vector field $X:TQ\rightarrow TTQ$ on $TQ$ is said to
be integrable if for any $v\in TQ$ the vector $X_{v}\in T_{v}TQ$
is integrable. This means that: 
\begin{equation}
T\tau_{Q}\circ X=\tau_{TQ}\circ X.\label{EQ4}
\end{equation}
Obviously, $T\tau_{Q}:TTQ\rightarrow TQ$ is the tangent mapping of
$TTQ$ onto $TQ$ and $X$ is defined with a mapping (cross-section)
of $TQ$ into $TTQ$. This is an absolute geometric language \cite{AbrMar78,Mac68,JJS_87,JJS91},
but everything becomes clear in the local coordinate systems $\left(q^{i},v^{i}\right)$.
Then any integrable vector field is expressed locally as 
\begin{equation}
X=v^{i}\frac{\partial}{\partial q^{i}}+\Phi^{i}\left(q^{j},v^{j}\right)\frac{\partial}{\partial v^{i}}.\label{EQ5}
\end{equation}
The corresponding integral curves satisfy the following local equations:
\begin{equation}
\frac{dq^{i}}{dt}=v^{i},\qquad\frac{dv^{i}}{dt}=\Phi^{i}\left(q^{j},v^{j}\right),\label{EQ6}
\end{equation}
i.e., briefly: 
\begin{equation}
\frac{d^{2}q^{i}}{dt^{2}}=\Phi^{i}\left(q^{j},\frac{dq^{j}}{dt}\right).\label{EQ7}
\end{equation}
This is the general form of Newtonian equations. It is seen that this
structure is inherent in the tangent bundle $TQ$. Nevertheless, it
is too poor for developing the details of both unconstrained and constrained
dynamics. Basing merely on those elements we are unable to define
such important concepts like energy, power, and work, necessary for
dealing with constraints in either d'Alembert or vakonomic sense.
The missing elements are those in a sense contained in another concept
of the state space, namely the cotangent bundle $T^{*}Q$, i.e., the
set-theoretical union of all spaces $T_{q}^{*}Q$ dual to $T_{q}Q$,
\begin{equation}
T^{*}Q=\underset{q\in Q}{\bigcup}T_{q}^{*}Q.\label{EQ8}
\end{equation}
The natural projection from this bundle onto $Q$ will be denoted
by $\tau_{Q}^{*}$, 
\begin{equation}
\tau_{Q}^{*}:T^{*}Q\rightarrow Q,\qquad\tau_{Q}^{*}\left(T_{q}^{*}Q\right)=\left\{ q\right\} .\label{EQ9}
\end{equation}
This cotangent bundle, i.e., Hamiltonian state space of the system,
dual to the Newton state space $TQ$, is endowed with the canonical
symplectic geometry \cite{AbrMar78,Mac68,JJS_87} based on the two-form:
\begin{equation}
\gamma=d\theta,\label{EQ10}
\end{equation}
where $\theta$ is the canonical Cartan differential one-form: 
\begin{equation}
\theta_{p}:=p\circ T\left.\tau_{Q}^{*}\right._{p}.\label{EQ11}
\end{equation}
Introducing components $p_{i}$ of $p$ we obtain coordinate systems
$\left(q^{i},p_{i}\right)$ defined locally on $T^{*}Q$. It is well
known and can be easily checked that locally 
\begin{equation}
\theta=p_{i}dq^{i},\qquad\gamma=d\theta=dp_{i}\wedge dq^{i}.\label{EQ12}
\end{equation}
This is another version of the state space. Its peculiarity is the
existence of the tensor fields (\ref{EQ12}). Unlike this, there is
no natural lift of curves from $Q$ to $T^{*}Q$, and the elements
of fibres, i.e., canonical momenta $p\in T^{*}Q$ are not operationally
interpretable. Nevertheless, it is a peculiarity of analytical mechanics,
especially with constraints, that the both state spaces $TQ$ and
$T^{*}Q$, i.e., the $\mu$-phase space and the usual (Hamiltonian)
state space, must be used. The correspondence between them is based
on the use of Lagrangian or Hamiltonian. Namely, the Lagrange function
$L:TQ\rightarrow\mathbb{R}$ gives rise to the Legendre transformation
$\mathcal{L}:TQ\rightarrow T^{*}Q$, namely 
\begin{equation}
\mathcal{L}(v):=D_{v}\left(L|T_{q}Q\right),\qquad v\in T_{q}Q.\label{EQ13}
\end{equation}
Analytically: 
\begin{equation}
\mathcal{L}:\left(q^{i},v^{i}\right)\rightarrow\left(q^{i},p_{i}\right)=\left(q^{i},\frac{\partial L}{\partial v^{i}}\right).\label{EQ14}
\end{equation}
The energy function $E:TQ\rightarrow\mathbb{R}$ is given by: 
\begin{equation}
E(v)=\left\langle \mathcal{L}(v),v\right\rangle -L(v)=v^{i}\frac{\partial L}{\partial v^{i}}-L(v).\label{EQ15}
\end{equation}
If $\mathcal{L}$ is invertible, the necessary although non-sufficient
condition for this is that the Hessian of $L$ is non-vanishing, 
\begin{equation}
\det\left[\frac{\partial^{2}L}{\partial v^{i}\partial v^{j}}\right]\neq0,\label{EQ16}
\end{equation}
then $E$ gives rise to the Hamilton function $H:T^{*}Q\rightarrow\mathbb{R}$,
namely: 
\begin{equation}
E=H\circ\mathcal{L},\qquad H=E\circ\mathcal{L}^{-1}.\label{EQ17}
\end{equation}
In this paper it is always assumed.

Let us only mention incidentally that if $\mathcal{L}\left(TQ\right)=M\subset T^{*}Q$,
the Hamilton function is still defined, but only on $M$, or more
rigorously, on the connected components of $M$. Although this situation
is also physically interesting, we do not consider it here \cite{AbrMar78,JJS91}.
Let us mention finally, that in the hyper-regular case, when $\mathcal{L}$
is invertible, the inverse mapping is analytically given by the following
expression analogous to (\ref{EQ14}): 
\begin{equation}
\mathcal{L}^{-1}:\left(q^{i},p_{i}\right)\rightarrow\left(q^{i},v^{i}\right)=\left(q^{i},\frac{\partial H}{\partial p_{i}}\right).\label{EQ18}
\end{equation}
The variational principle for the Lagrangian dynamical system, 
\begin{equation}
\delta\int Ldt=0,\label{EQ19}
\end{equation}
leads to the usual Lagrange equations of the second kind: 
\begin{equation}
-\frac{\delta L}{\delta q^{i}(t)}=\frac{d}{dt}\frac{\partial L}{\partial\dot{q}^{i}}-\frac{\partial L}{\partial q^{i}}=0,\qquad i=1,\ldots,n.\label{EQ20}
\end{equation}
If in addition to variational forces derivable from $L$ also some
other, first of all dissipative ones $D_{i}$, are present, then equations
(\ref{EQ20}) are replaced by: 
\begin{equation}
\frac{d}{dt}\frac{\partial L}{\partial\dot{q}^{i}}-\frac{\partial L}{\partial q^{i}}=D_{i}.\label{EQ21}
\end{equation}
Let us stress that $D_{i}$ geometrically is a covector field. The
energy balance has the form: 
\begin{equation}
\frac{dE}{dt}=-\frac{\partial L}{\partial t}+D_{i}\frac{dq^{i}}{dt}.\label{EQ22}
\end{equation}
The non-conservation of $E$ follows from two circumstances: 
\begin{itemize}
\item 1) the explicit dependence of $L$ on $t$ (in addition to usual dependence
of $q^{i}$, $\dot{q}^{i}$ on $t$), 
\item 2) the power of non-variational forces equals $D_{i}\left(dq^{i}/dt\right)$;
$D_{i}$ is a covariant vector. 
\end{itemize}
This was description of unconstrained systems. Let us now assume that
the system is subject to holonomic constraints which restrict its
motion to some submanifold $W\subset Q$. One assumes then that also
the tangent bundle, i.e., manifold of Newtonian states is restricted
from $TQ$ to $TW$. Nevertheless, some qualitative comments are necessary
here. First of all: what does it mean that the motion is restricted
to $W$ and what is the status of resulting equations of motion? First
of all let us assume some metric tensor $g$ in $Q$ and some field
of forces which on $W$ is tangent to it. They may be encoded in the
expression for the total potential energy. Let $W$ be given by equations
\begin{equation}
F_{a}\left(q^{1},\ldots,q^{n}\right)=0,\qquad a=1,\ldots,m,\label{EQ23}
\end{equation}
where 
\begin{equation}
dF_{1}(q)\wedge dF_{2}(q)\wedge\ldots\wedge dF_{a}(q)\neq0\label{EQ24}
\end{equation}
for any $q\in W$, i.e., satisfying (\ref{EQ23}). We assume that
the potential of forces $g$-orthonormal to $W$ is given by 
\begin{equation}
U_{\kappa}=\frac{1}{2}\kappa^{ab}F_{a}F_{b},\label{EQ25}
\end{equation}
and the total potential energy has the form: 
\begin{equation}
U=U_{\kappa}+V,\label{EQ26}
\end{equation}
where $V$ is independent on the quantity $\kappa$.

The coefficients $\kappa^{ab}$ are assumed to be sufficiently large
to force the system to remain permanently in a small neighbourhood
of $W$. Equations of motion following from the Lagrangian $L=T-U$
are dependent on the parameters $\kappa^{ab}$ and one can perform
the limit transition $\kappa^{ab}\rightarrow\infty$. One can show
\cite{Arnold_97} that in the limit, with initial conditions compatible
with $W$, one obtains that the trajectory is constrained in $W$
and satisfies the following equations: 
\begin{equation}
\frac{d}{dt}\frac{\partial L^{W}}{\partial\dot{x}^{\mu}}-\frac{\partial L^{W}}{\partial x^{\mu}}=0,\qquad\mu=1,\ldots,(n-m),\label{EQ27}
\end{equation}
where $L^{W}$ is the restriction of $L:TQ\rightarrow\mathbb{R}$
to the tangent submanifold $TW\subset TQ$. The quantities $x^{\mu}$
are local coordinates (parameters) on $W$; $q^{i}$ are their functions:
\begin{equation}
q^{i}=f^{i}\left(x^{\mu}\right).\label{EQ28}
\end{equation}
Obviously, we have 
\begin{equation}
L^{W}\left(x,\dot{x}\right):=L\left(f^{i}(x),\frac{\partial f^{j}}{\partial x^{\mu}}\dot{x}^{\mu}\right).\label{EQ29}
\end{equation}
One can show that the formula similar to (\ref{EQ21}), (\ref{EQ27})
is also satisfied when dissipative forces are present: 
\begin{equation}
\frac{d}{dt}\frac{\partial L^{W}}{\partial\dot{x}^{\mu}}-\frac{\partial L^{W}}{\partial x^{\mu}}=D^{W}\!_{\mu},\label{EQ30}
\end{equation}
where the covariant dissipative force on $W$ is given by 
\begin{equation}
D^{W}\!_{\mu}\left(x,\dot{x}\right)=D_{i}\left(f^{j}(x),\frac{\partial f^{l}}{\partial x^{\lambda}}\dot{x}^{\lambda}\right)\frac{\partial f^{i}}{\partial x^{\mu}}.\label{EQ31}
\end{equation}
To discuss and understand properly the peculiarity of vakonomic systems
it is however more natural to remain on the level of the implicite
description. Namely, when dealing with a holonomically constrained
dynamical system, the principle of the action minimum (or in any case,
stationary point) (\ref{EQ19}) is replaced by that of conditional,
i.e., constrained extremum (stationary point): 
\begin{equation}
\delta\int Ldt=0,\qquad F_{a}(q)=0,\qquad a=1,\ldots,m.\label{EQ32}
\end{equation}
It claims that realistic motions are those for which the variational
condition in (\ref{EQ32}) is satisfied for those virtual motions
which are compatible with the second subsystem. But, according to
the Lusternik theorem about conditional extrema, they are those which
give stationary value to the functional: 
\begin{equation}
\int\left(L+\mu^{a}F_{a}\right)dt=\int L^{\prime}dt,\label{EQ33}
\end{equation}
where $\mu^{a}$ are apriori non-specified Lagrange multipliers, in
a sense additional state variables. Performing the variational procedure
for all variables we obtain: 
\begin{equation}
\frac{d}{dt}\frac{\partial L}{\partial\dot{q}^{i}}-\frac{\partial L}{\partial q^{i}}=\mu^{a}\frac{\partial F_{a}}{\partial q^{i}},\qquad F_{a}(q)=0.\label{EQ34}
\end{equation}
And similarly for systems with dissipative forces we obtain: 
\begin{equation}
\frac{d}{dt}\frac{\partial L}{\partial\dot{q}^{i}}-\frac{\partial L}{\partial q^{i}}=D_{i}+R_{i}=D_{i}+\mu^{a}\frac{\partial F_{a}}{\partial q^{i}},\qquad F_{a}(q)=0.\label{EQ35}
\end{equation}
The quantities $R_{i}$ given by 
\begin{equation}
R_{i}=\mu^{a}\frac{\partial F_{a}}{\partial q^{i}}\label{EQ36}
\end{equation}
are reaction forces responsible for maintaining the constraints given
by (\ref{EQ23}). The total system (\ref{EQ35}) is imposed on the
time-dependence of the $(n+m)$ quantities $q^{i}$, $\mu^{a}$.

Let us notice some essential point. Reaction forces $R_{i}$ are defined
and (in general) non-vanishing on the manifolds of constrained motion
$W\subset Q$, $TW\subset TQ$. Unlike this, the physical reactions
of approximately constrained motion exactly vanish on $W$ and in
a small neighbourhood of $W$ they are very large and act attractively
towards $W$. Their potential is just $U_{\kappa}$ (\ref{EQ25}),
(\ref{EQ26}). Because of this, for any finite $\kappa^{ab}$ the
system performs quick oscillations about $W$, with a very small amplitude.
This difference between physical reactions for finite $\kappa^{ab}$
and the ideal reactions (\ref{EQ36}) is very essential when the quantized
problem is investigated. Namely, the mentioned small oscillations
about $W$ are then labelled by large quantum numbers.

\section{Quantization}

The problem becomes complicated and in certain situations the results
may by qualitatively different from those obtained by the intrinsic
quantization in $W$. We do not get into details here. We only mention
what is meant here by the ``intrinsic quantization''. Let the metric
tensor on $W$ induced by the restriction of $g$ to $W$ be denoted
by $g^{W}$, just the pull-back of $g$, 
\begin{equation}
g^{W}=\imath^{*}\cdot g,\label{EQ37}
\end{equation}
where $\imath:W\rightarrow Q$ is the natural injection, and let the
effective potential energy on $W$ be denoted by $V^{W}$, 
\[
V^{W}=\imath^{*}\cdot V=V\circ\imath.
\]
Then the intrinsic $W$-quantization is based on the Hamilton operator:
\begin{equation}
\mathbb{H}=-\frac{\hbar^{2}}{2m}\Delta\left[g^{W}\right]+V^{W},\label{EQ38}
\end{equation}
where $\Delta\left[g^{W}\right]$ is the Laplace-Beltrami operator
acting on wave functions on $W$, 
\begin{equation}
\Delta\left[g^{W}\right]\Psi=\frac{1}{\sqrt{\left|g^{W}\right|}}\underset{\mu,\nu}{\sum}\frac{\partial}{\partial x^{\mu}}\left(\sqrt{\left|g^{W}\right|}g^{W\mu\nu}\frac{\partial\Psi}{\partial x^{\nu}}\right),\label{EQ39}
\end{equation}
and obviously 
\begin{equation}
\left|g^{W}\right|=\sqrt{\left|\det\left[g_{\mu\nu}^{W}\right]\right|}.\label{EQ40}
\end{equation}
It is clear that the quantum numbers of $\mathbb{H}$ may happen to
interfere badly with those of the $U_{\kappa}$-terms (\ref{EQ25}),
(\ref{EQ26}) for large but finite $\kappa$. This may be interesting
in quantum studies of the constrained motion, and really it seems
that there are some doubful problems there. However, here we do not
deal with those problems there. In any case it seems that there are
two not necessarilly equivalent approaches: one based on the formal
restriction of the metric to constraints, and another, based on the
properties of the potential. In any case even here there are some
indications that the d'Alembert procedure need not by convincing.

\section{Description of dynamics by d'Alembert and vakonomic theories}

Let us go back to classical problems. It is clear that the reactions
(\ref{EQ36}) do no work along any curve compatible with constraints,
their power vanishes then: 
\begin{equation}
R_{i}\frac{dq^{i}}{dt}=\mu^{a}\frac{\partial F_{a}}{\partial q^{i}}\frac{dq^{i}}{dt}=0\label{EQ41}
\end{equation}
if for any $t$, $q^{i}(t)\in W$; this is a direct consequence of
(\ref{EQ32}). In classical textbooks it is formulated as the principle
of vanishing work along any trajectory compatible with constraints,
\begin{equation}
R_{i}\delta q^{i}=\mu^{a}\frac{\partial F_{a}}{\partial q^{i}}\delta q^{i}=0\label{EQ42}
\end{equation}
if $\delta q^{i}$ is tangent to $W$, i.e., if 
\begin{equation}
\frac{\partial F_{a}}{\partial q^{i}}\delta q^{i}=0\label{EQ43}
\end{equation}
quite independently of any dynamical equations. In XIX-XX-th centuries
the analysis of virtual displacements was treated very seriously and
some art concerning the constitutive relations for them was developed.
Let us remind the Appell-Chetaev, Gauss and other treatments. They
were motivated by non-holonomic constrains, in particular for ones
non-linear in velocities. Some of them were very inventive, although
there are also opinions that certain artefacts and misconceptions
appeared in them \cite{NeiFuf81,JJS_87}. The point is that for Lusternik
procedures one obtains different, non-convincing results.

Let us assume that the system is subject to some purely non-holonomic
constraints $M\subset TQ$ linear in velocities. This means that $\tau_{Q}(M)=Q$
and for any $q\in Q$ the manifold $M_{q}:=M\cap T_{q}Q$ is a linear
$(n-m)$-dimensional subspace. One can also admit it to be an affine
subspace of $T_{q}Q$, i.e., some translation of a linear subspace
by a vector non-contained in $M_{q}$. However, for simplicity we
do not consider this modification here. Therefore, $M$ has equations
of the form: 
\begin{equation}
F_{a}(q,v)=\omega_{ai}(q)v^{i}=0,\qquad a=1,\ldots,m.\label{EQ44}
\end{equation}
We assume also that the constraints are not semi-holonomic, i.e.,
that the Pfaff systems 
\begin{equation}
\omega_{a}=\omega_{ai}(q)dq^{i}=0\label{EQ45}
\end{equation}
is not maximally integrable. If it was, the manifold $Q$ would be
foliated by the $m$-dimensional family of mutually disjoint $(n-m)$-dimensional
integral surfaces of (\ref{EQ45}). This would be the case if the
following condition was satisfied: 
\begin{equation}
d\omega_{a}\wedge\omega_{1}\wedge\ldots\wedge\omega_{m}=0,\qquad a=1,\dots,m.\label{EQ46}
\end{equation}

So, we assume that the problem is purely non-holonomic, does not reduce
to the family of holonomic constraints, and therefore, the family
of configurations which may be reached from any point of $Q$ is $n$-dimensional.
The statement ``may be reached'' is meant here in a purely geometric,
non-dynamical sense: just reached by any integral curve of (\ref{EQ45}),
without any dynamical assumption. Let us again apply the Lusternik
theorem to the constrained variational problem: 
\begin{equation}
\delta\int L\left(q(t),\dot{q}(t)\right)dt=0,\qquad F_{a}\left(q(t),\dot{q}(t)\right)=\omega_{ai}(q)\frac{dq^{i}}{dt}=0.\label{EQ47}
\end{equation}

So, we look for the extremals (more precisely-stationary points) of
the functional 
\begin{equation}
\left[q(\cdot)\right]\rightarrow\int L\left(q(t),\dot{q}(t)\right)dt\label{EQ48}
\end{equation}
on the family of curves satisfying the second condition of (\ref{EQ47}).
But this means that again we introduce additional parameters $\mu^{a},\, a=1,\dots,m$,
i.e., Lagrange multipliers, and look for the family of unconstrained
dynamical systems for $t\rightarrow\left(q^{i}(t),\mu^{a}(t)\right)$
described by: 
\begin{equation}
\delta\int\widetilde{L}\left(q(t),\dot{q}(t);\mu(t)\right)dt=0,\label{EQ49}
\end{equation}
where $\widetilde{L}$ is given by: 
\begin{equation}
\widetilde{L}\left(q(t),\dot{q}(t);\mu(t)\right)=L\left(q(t),\dot{q}(t)\right)+\mu^{a}(t)F_{a}\left(q(t),\dot{q}(t)\right).\label{EQ50}
\end{equation}
Performing carefully the unconstrained variational procedure for $\left(q(t),\mu(t)\right)$,
we finally obtain the system: 
\begin{equation}
\frac{d}{dt}\frac{\partial L}{\partial\dot{q}^{i}}-\frac{\partial L}{\partial q^{i}}=R_{i},\qquad\omega_{ai}(q)\frac{dq^{i}}{dt}=0,\label{EQ51}
\end{equation}
where 
\begin{equation}
R_{i}=-\frac{d\mu^{a}}{dt}\omega_{ai}+\mu^{a}\left(\frac{\partial\omega_{aj}}{\partial q^{i}}-\frac{\partial\omega_{ai}}{\partial q^{j}}\right)\frac{dq^{j}}{dt}.\label{EQ52}
\end{equation}
Similarly, when non-variational interactions are admitted, we have:
\begin{equation}
\frac{d}{dt}\frac{\partial L}{\partial\dot{q}^{i}}-\frac{\partial L}{\partial q^{i}}=D_{i}+R_{i},\label{EQ53}
\end{equation}
where $R_{i}$ is as above in (\ref{EQ52}) and $D_{i}$ are covariant
components of forces non-derivable from Lagrangian. It is seen that
the structure of $R_{i}$ differs form (\ref{EQ36}) by the ``magnetic-like''
second term of (\ref{EQ52}). But nevertheless one can think just
like in holonomic systems and assume that reactions do no work on
virtual displacements compatible with constraints. Therefore, they
are power-free on velocities satisfying second equation of (\ref{EQ51}),
i.e., 
\begin{equation}
R_{i}\frac{dq^{i}}{dt}=0,\qquad\textrm{if}\qquad\omega_{ai}\frac{dq^{i}}{dt}=0.\label{EQ54}
\end{equation}
Then 
\begin{equation}
R_{i}=\lambda^{a}\omega_{ai}=-\frac{d\mu^{a}}{dt}\omega_{ai}.\label{EQ55}
\end{equation}
Therefore, again we would have 
\begin{equation}
\frac{d}{dt}\frac{\partial L}{\partial\dot{q}^{i}}-\frac{\partial L}{\partial q^{i}}=D_{i}+\lambda^{a}\omega_{ai}\label{EQ56}
\end{equation}
just like in (\ref{EQ35}), (\ref{EQ36}), this time with $\omega_{ai}$
instead of $\partial F_{a}/\partial q^{i}$. Moreover, it turns out
that (\ref{EQ56}) is not only geometrically possible but just physically
correct for all known problems of sliding-free rolling on rough surfaces.
But mathematically (\ref{EQ56}) are non-variational in structure.
An explanation was that the corresponding reactions arise through
the non-variational friction phenomena \cite{NeiFuf81}.

Nevertheless the variational equations (\ref{EQ51})--(\ref{EQ53})
with their intriguing ``magnetic-like'' correction terms seemed
to be so exciting that they were intensively studied even for the
purely mathematical/geometrical purposes. Various names were used
for them like vakonomy (from ``variational axiomatic kind'').

It is clear that the second term in (\ref{EQ52}) does not influence
the energy balance of reactions, 
\begin{equation}
R_{i}\frac{dq^{i}}{dt}=-\frac{d\mu^{a}}{dt}\omega_{ai}\frac{dq^{i}}{dt}.\label{EQ57}
\end{equation}

It is admissible due to its very ``magnetic'' structure. In the
period when the analysis of various definitions of ``virtual displacements''
was fashionable, there were some attempts to modify in a non-variational
way the mutual ratio of coefficients at the two terms of (\ref{EQ52}).
But the violation of the variational character of (\ref{EQ52}) seems
to be the price paid for nothing. Besides, in non-holonomic systems
the reaction of the classical expression for the virtual displacement,
\begin{equation}
\omega_{ai}(q)\delta q^{i}=0,\label{EQ58}
\end{equation}
although physically correct in the rolling case, seems to be frustrating.
Indeed, when the higher-order terms are neglected, then 
\begin{equation}
\delta F_{a}=F_{a}\left(q+\delta q\right)-F_{a}\left(q\right)=\frac{d}{dt}\left(\omega_{ai}\frac{dq^{i}}{dt}\right)+\left(\omega_{aj,i}-\omega_{ai,j}\right)\frac{dq^{j}}{dt}\delta q^{i}.\label{EQ59}
\end{equation}
But (\ref{EQ58}) does not imply $\delta F_{a}$ to vanish. Instead,
the two terms in (\ref{EQ59}) should separately vanish. But those
$2m$ conditions would be too much. The only possibility to remain
within variational scheme is just to take only $m$ conditions: 
\begin{equation}
\frac{d}{dt}\left(\omega_{ai}\dot{q}^{i}\right)+\left(\omega_{aj,i}-\omega_{ai,j}\right)\dot{q}^{j}\delta q^{i}=0.\label{EQ60}
\end{equation}

Yes, but as said above, in the physical case of sliding-free rolling,
one uses (\ref{EQ56}), (\ref{EQ58}). What would be the domain of
applications of incomparatively more elegant vakonomic systems? First
of all, let us notice that the VAK-dynamics is very interesting from
the purely mathematical point of view and that it has given rise to
the completely new domain of mathematical physics based on variational
principles \cite{Be-Die_05,deLeon_03,JJS_87}. What concerns doubtful
practical applications of vakonomy, they are in financial dynamical
problems. And it is not clear if they should not be used in molecular
and nuclear dynamics \cite{JJS_87}. But what is very important, it
seems that the vakonomic ideas may be very useful in certain problems
of the control theory. We mean here the active control procedures,
not the friction-based methods. There is a feeling that the active
problems are much more effective and natural due to their invariance
properties. And besides, other methods are completely non-useful in
nonlinear and differential higher-order problems.

Let us begin with the nonlinear case. We assume that the manifold
of admissible virtual velocities at the configuration $q\in Q$ is
given by the set $V_{q}$. For simplicity we assume that at any $q\in Q$
all the manifolds $V_{q}$ have the same dimension $(n-m)$. Therefore,
the total manifold of constraints $M$, 
\begin{equation}
M=\underset{q\in Q}{\bigcup}V_{q},\label{EQ61}
\end{equation}
is a geometric plane given by equations: 
\begin{equation}
F_{a}\left(q^{1},\dots,q^{n};v^{1},\dots,v^{n}\right)=0,\qquad a=1,\dots,m.\label{EQ62}
\end{equation}

For simplicity we assume that the range of positions $\left(q^{1},\dots,q^{n}\right)$
is non-restricted and it is really the range of velocity variables
that is subject to constraints. Therefore, 
\begin{equation}
Rank\left[\frac{\partial F_{a}}{\partial v^{i}}\right]=n-m.\label{EQ63}
\end{equation}
When the constraints (\ref{EQ61}), (\ref{EQ62}) are nonlinear, then
the procedure of eliminating constraints with the help of the above-quoted
procedures is literally meaningless. Indeed, in general the principle
of virtual displacements does not allow us to formulate true equations
of motion. In fact, when $V_{q}$ is a general differential sub-manifold
of $T_{q}Q$, then the linear shell (closure) of $V_{q}$-elements
will have higher dimension than $V_{q}$ and in particular, it may
coincide with $T_{q}Q$ itself. There were many attempts to prevent
this. In a sense, the simplest and most general of them was the Appell-Chetaev
procedure \cite{Gut_71,NeiFuf81}. Let us quote the general ideas
of that methods. Namely, we can try to go back to the philosophy of
virtual displacements and define them as ones satisfying: 
\begin{equation}
\frac{\partial F_{a}}{\partial\dot{q}^{i}}\delta q^{i}=0,\qquad a=1,\dots,m,\label{EQ64}
\end{equation}
again with the summation convention applied. This is equivalent to
\begin{equation}
R(q,v)_{i}=\lambda^{a}\frac{\partial F_{a}}{\partial v^{i}}\label{EQ65}
\end{equation}
for reaction forces. This follows from the fact that for any $q\in Q$,
$v\in V_{q}$, and $w\in T_{v}M_{q}\subset T_{q}Q$ the following
equation holds: 
\begin{equation}
\left\langle R\left(q,v\right),w\right\rangle =R\left(q,v\right)_{i}w^{i}=0.\label{EQ66}
\end{equation}
Therefore, equations of motion may be written in the variational form:
\begin{equation}
\left(\frac{d}{dt}\frac{\partial L}{\partial\dot{q}^{i}}-\frac{\partial L}{\partial q^{i}}\right)\delta q^{i}=0,\qquad F_{a}\left(q,\dot{q}\right)=0,\qquad\frac{\partial F_{a}}{\partial\dot{q}^{i}}\delta q^{i}=0,\label{EQ67}
\end{equation}
or in the following explicit form: 
\begin{equation}
\frac{d}{dt}\frac{\partial L}{\partial\dot{q}^{i}}-\frac{\partial L}{\partial q^{i}}=D_{i}+\lambda^{a}\frac{\partial F_{a}}{\partial\dot{q}^{i}},\qquad F_{a}\left(q,\dot{q}\right)=0.\label{EQ68}
\end{equation}
These equations (\ref{EQ68}) must be jointly solved for the time
dependence of $\left(q^{i},\lambda^{a}\right)$, in general without
the possibility of separating $\lambda^{a}$ and substituting them
to the equations for $q^{i}$.

Let us stress that the above procedure defining $\delta q^{i}$ is
mathematically correct and geometric, although in general physically
non-convincing. It does not seem to refer to the energy balance or
mechanical work. It also fails to describe the tangent vector in the
functional manifold of motions compatible with constrains. In general,
the XIX-th century procedure for establishing some explicit rules
for virtual displacements, in spite of its inventive power, seems
to be rather misleading from the point of view of the active control
problems of servomechanisms. In particular, this concerns the problems
of motion of satellites and other space-moving objects. It seems that
the d'Alembert procedure and the acceleration-dependent Gauss method
are definitely less reliable in such problems.

This seems particularly hopeful when we are looking for the program
forces which keep velocity constant or appropriately programmed, both
in the sense of direction or magnitude. Let us consider a general
mechanical system with nonlinear non-holonomic constraints. We begin
with first-order differential constraints, i.e., ones imposed on configurations
and generalized velocities, but now without any assumption of linearity.
Let us begin with the variational principle 
\begin{equation}
\delta\int L\left(q(t),\dot{q}(t)\right)dt=0\label{EQ69}
\end{equation}
with extra imposed nonlinear constraints: 
\begin{equation}
F_{a}\left(q,\dot{q}\right)=0,\qquad a=1,\dots,m.\label{EQ70}
\end{equation}
The resulting equations of motion have the form following from the
Lusternik principle: 
\begin{equation}
\frac{d}{dt}\frac{\partial L}{\partial\dot{q}^{i}}-\frac{\partial L}{\partial q^{i}}=R_{i},\qquad F_{a}\left(q,\dot{q}\right)=0,\label{EQ71}
\end{equation}
where the variational reactions are given by: 
\begin{equation}
R_{i}=\lambda^{a}\frac{\partial F_{a}}{\partial q^{i}}-\frac{d}{dt}\left(\lambda^{a}\frac{\partial F_{a}}{\partial\dot{q}^{i}}\right)=-\frac{d\lambda^{a}}{dt}\frac{\partial F_{a}}{\partial\dot{q}^{i}}+\lambda^{a}\frac{\partial F_{a}}{\partial q^{i}}-\lambda^{a}\frac{\partial^{2}F_{a}}{\partial\dot{q}^{i}\partial\dot{q}^{j}}\frac{d^{2}q^{j}}{dt^{2}}-\lambda^{a}\frac{\partial^{2}F_{a}}{\partial\dot{q}^{i}\partial q^{j}}\frac{dq^{j}}{dt}.\label{EQ72}
\end{equation}
When dissipative forces are present, then (\ref{EQ71}) becomes: 
\begin{equation}
\frac{d}{dt}\frac{\partial L}{\partial\dot{q}^{i}}-\frac{\partial L}{\partial q^{i}}=D_{i}+R_{i},\qquad F_{a}\left(q,\dot{q}\right)=0.\label{EQ73}
\end{equation}

Let us observe that now the quantities $\left(q^{i},\lambda^{a}\right)$
occur in equations of motion on the almost equal footing as a kind
of generalized coordinates. Because of this now there is really no
chance for elimination of reactions $\lambda^{a}$ from our equations.
They must be found simultaneously from those equations.

It is clear that $R_{i}$ in (\ref{EQ72}) contains also the Appell-Chetaev
term 
\begin{equation}
-\frac{d\lambda^{a}}{dt}\frac{\partial F_{a}}{\partial\dot{q}^{i}},
\end{equation}
as mentioned above, geometrically well defined. Nevertheless, there
are also three other terms, much more interesting from the point of
view of the active control. For example, we may be interested in fixing
with the help of some constraints the absolute value of velocity of
some space-moving object. It is clear that the d'Alembert procedure
developed for non-holonomic constraints with equations linear in velocities
rather completely fails when dealing with such problems. Although,
one must say that there were some attempts of using them, based on
a strange procedure of the limit transition connected with the reduction
of the number of degrees of freedom. But they were non-convincing
and subject to certain criticism \cite{NeiFuf81}. In any case it
seems obvious that the procedure of the friction-based explanation
of control seems to be exotic there. Let us compare the two procedures.
We assume the simplest Lagrangian of the from: 
\begin{equation}
L=\frac{m}{2}g_{ij}\dot{q}^{i}\dot{q}^{j}-V(q),\label{EQ74}
\end{equation}
where the metric tensor $g_{ij}$ is assumed to have constant coefficients
(so, it is curvature-free). Nonlinear constraints fix the absolute
value of velocity, 
\begin{equation}
F\left(\dot{q}\right)=g_{ij}\dot{q}^{i}\dot{q}^{j}-c^{2},\label{EQ75}
\end{equation}
where $c$ is a constant. The resulting Appell-Chetaev leads to the
following system of equations: 
\begin{equation}
m\frac{d^{2}q^{i}}{dt^{2}}-\lambda\frac{dq^{i}}{dt}+g^{ij}\frac{\partial V}{\partial q^{j}}=0,\qquad g_{ij}\frac{dq^{i}}{dt}\frac{dq^{j}}{dt}-c^{2}=0.\label{EQ76}
\end{equation}
Unlike this, the variational vakonomic procedure gives us: 
\begin{equation}
\left(m+\lambda\right)\frac{d^{2}q^{i}}{dt^{2}}-\frac{d\lambda}{dt}\frac{dq^{i}}{dt}+g^{ij}\frac{\partial V}{\partial q^{j}}=0,\qquad g_{ij}\frac{dq^{i}}{dt}\frac{dq^{j}}{dt}-c^{2}=0.\label{EQ77}
\end{equation}
It is clear that the both systems are different and non-equivalent.
Again in the vakonomic Lusternik procedure $\lambda$ is a kind of
degree of freedom subject jointly with $q^{i}$ to a system of differential
equations. And really it seems more natural here to use $\lambda$
as the active control factor. Besides, it influences the inertial
properties (mass) of the body by introducing the additional, dynamical
mass term $\lambda\left(d^{2}q^{i}/dt^{2}\right)$.

It is interesting to interpret physically some terms of (\ref{EQ72}).
The third term may be interpreted as a channel which controls the
inertia, i.e., the effective mass of the body. The resulting control
may regulate, e.g., stabilize the angular velocity of rotors. The
fourth term seems to influence the damping and acceleration forces,
and also some gyroscopic or magnetic-like behavior, especially when
combined appropriately with the second term.

Let us stress that we are thinking here about the problems of the
active control. This brings about the question if our reaction are
adiabatic, i.e., if they do no work. As expected, it turns out that
in general it is not the case. Namely, one can show that the power
of vakonomic reactions need not vanish. Indeed, 
\begin{equation}
R_{i}\frac{dq^{i}}{dt}=\lambda^{a}\frac{dF_{a}}{dt}-\frac{d}{dt}\left(\lambda^{a}\frac{\partial F_{a}}{\partial\dot{q}^{i}}\dot{q}^{i}\right).\label{EQ78}
\end{equation}
Obviously, the first term vanishes in a consequence of the very equations
of constraints. Therefore, 
\begin{equation}
R_{i}\dot{q}^{i}=-\frac{d}{dt}\left(\lambda^{a}\frac{\partial F_{a}}{\partial\dot{q}^{i}}\dot{q}^{i}\right).\label{EQ79}
\end{equation}
And in general case this is non-vanishing. But it is interesting to
ask for the special case when that term does vanish. It turns out
that it happens when constraints are linear in velocities. Indeed,
linear functions are homogeneous of degree one, i.e., the following
holds for them: 
\begin{equation}
v^{i}\frac{\partial F_{a}}{\partial v^{i}}=F_{a}.\label{EQ80}
\end{equation}
Therefore, (\ref{EQ79}) becomes: 
\begin{equation}
R_{i}\dot{q}^{i}=-\frac{d}{dt}\left(\lambda^{a}F_{a}\right).\label{EQ81}
\end{equation}
This expression evidently vanishes along any curve compatible with
the imposed constraints. It is even more, namely (\ref{EQ81}) vanishes
when functions $F_{a}$ satisfy the following differential equation:
\begin{equation}
v^{i}\frac{\partial F_{a}}{\partial v^{i}}=F_{a},\label{EQ82}
\end{equation}
which implies that (\ref{EQ79}) becomes 
\begin{equation}
\frac{d}{dt}\left(\lambda^{a}F_{a}\right)=0\label{EQ83}
\end{equation}
along any curve compatible with constraints. Moreover, expression
(\ref{EQ79}) vanishes also if the following holds: 
\begin{equation}
\left.v^{i}\frac{\partial F_{a}}{\partial v^{i}}\right|_{M}=0\label{EQ84}
\end{equation}
or, equivalently if for some functions $g_{a}\!^{b}$ well defined
in a neighbourhood of $M$ the equation 
\begin{equation}
v^{i}\frac{\partial F_{a}}{\partial v^{i}}=g_{a}\!^{b}F_{b}\label{EQ85}
\end{equation}
is satisfied. But those facts imply that the field of vectors $v^{i}\left(\partial/\partial v^{i}\right)$
is tangent to $M$. However, it is clear that this vector field is
a generator of the group of dilatations, namely $v\rightarrow e^{\tau}v$
in tangent spaces of $Q$. Therefore, its integral curves are half-spaces
of one-dimensional subspaces of the tangent spaces $T_{q}Q$. Or,
in the singular case, they are simply null elements of $T_{q}Q$.
This means that every $M$-manifold satisfying (\ref{EQ85}) is built
of one-dimensional subspaces of the tangent spaces $T_{q}Q$. And
every one-dimensional subspace, or rather half-subspace, is either
contained in $M$ or just disjoint with $M$. Therefore, the manifold
$M_{q}=M\cap T_{q}Q$ induces constraints on the directions in $T_{q}Q$,
but not on the magnitudes of vectors.

This implies an important feature of non-holonomic constraints: if
non-holonomic conditions $M$ restrict the manifold of directions
of velocities, but do not influence their magnitudes, then the Lusternik
reactions are adiabatic, i.e., they do no work. This result is interesting
and expected. It is difficult to realize the control, and in particular
stabilization of velocity, on the basis of the d'Alembert procedure,
without using the energy income to the system. Unlike this, it is
possible when we wish to keep or control the direction of motion.

Following (\ref{EQ79}) we can write the variational Lusternik energy
balance as follows: 
\begin{equation}
\frac{d}{dt}E\left[L,M\right]=\frac{d}{dt}\left(E[L]+\lambda^{a}\frac{\partial F_{a}}{\partial\dot{q}}\dot{q}^{i}\right)=D_{i}\dot{q}^{i}.\label{EQ86}
\end{equation}
Here we have obviously: 
\begin{equation}
E[L]=v^{i}\frac{\partial L}{\partial v^{i}}-L\label{EQ87}
\end{equation}
is the usual energy of mechanical system, whereas 
\begin{equation}
E[M]=\lambda^{a}\frac{\partial F_{a}}{\partial v^{i}}v^{i}\label{EQ88}
\end{equation}
is the energy of constraints alone.

As seen from (\ref{EQ86}), in the absence of dissipation, there is
a conservation of the total energy; the total means a combination
of $E[L]$ and $E[M]$. This conservation law follows also immediately
from the Lusternik dynamics.

Indeed, according to the general principles of mechanics, the total
energy of our system is: 
\begin{equation}
E\left[L[\lambda]\right]=\dot{q}^{i}\frac{\partial L[\lambda]}{\partial\dot{q}^{i}}-L[\lambda]=E[L]+\lambda^{a}\frac{\partial F_{a}}{\partial\dot{q}^{i}}\dot{q}^{i}-\lambda^{a}F_{a}.\label{EQ89}
\end{equation}
But on the $F_{a}$-constrained system the last term does vanish and
(\ref{EQ89}) is identical with $E[L,M]$. The additional term $\lambda^{a}\left(\partial F_{a}/\partial v^{i}\right)v^{i}$
describes the exchange of energy between the system of constraints
and our object. This is a rather reasonable process when controlling
the magnitude of velocity, in any case much more convincing than various
XIX-th century procedures based on more or less sophisticated inventions
of ``virtual displacements''. In any case it is so when discussing
the modern problems of the active time-dependent control agents.

The problem becomes much more essential when we admit non-holonomic
constraints depending also on higher-order time derivatives of generalized
coordinates, 
\begin{equation}
F_{A}\left(q,\dot{q},\ddot{q},\dots,\overset{(N)}{q}\right),\qquad A=1,\dots,m.\label{EQ90}
\end{equation}
Obviously, in this expression $\overset{(L)}{q}$ denotes the symbol
of the $L$-th-order derivatives. The variational Lusternik procedure
leads to the following system of equations of motion: 
\begin{equation}
\frac{d}{dt}\frac{\partial L}{\partial\dot{q}^{i}}-\frac{\partial L}{\partial q^{i}}=D_{i}+R_{i},\qquad F_{A}\left(q,\dot{q},\ddot{q},\dots,\overset{(N)}{q}\right)=0,\label{EQ91}
\end{equation}
where 
\begin{equation}
R_{i}=\overset{N}{\underset{L=0}{\sum}}(-1)^{L}\frac{d^{L}}{dt^{L}}\left(\lambda^{A}(t)\frac{\partial F_{A}}{\partial\overset{(L)}{q}^{i}}\right).\label{EQ92}
\end{equation}

It is a more academic, nevertheless also interesting problem what
are the equations of controlled motion of the system the uncontrolled
Lagrangian of which depends on higher-order derivatives, i.e., if
it has the general shape: 
\begin{equation}
L\left(q,\dot{q},\ddot{q},\dots,\overset{(M)}{q}\right)\label{EQ93}
\end{equation}
with the same meaning of symbols as in (\ref{EQ91}). Obviously, at
least formally the Lusternik procedure with constraints (\ref{EQ90})
is still formally applicable and leads to the following equations
of motion: 
\begin{equation}
\overset{N}{\underset{K=0}{\sum}}(-1)^{K+1}\frac{d^{K}}{dt^{K}}\frac{\partial L}{\partial\overset{(K)}{q}^{i}}=R_{i}\label{EQ94}
\end{equation}
again with $R_{i}$ given by (\ref{EQ92}). Principally in a similar
way we can discuss controlling constraints given by non-differential
expressions, e.g., integral or other, more general. This will be a
subject of the next paper. For some reasons maximally interesting
is the special case $N=2$, or also $N\geq2$. Then the reaction forces
are given by 
\[
R_{i}=\mu^{a}\frac{\partial F_{a}}{\partial q^{i}}-\frac{D}{Dt}\left(\mu^{a}\frac{\partial F_{a}}{\partial\dot{q}^{i}}\right)+\frac{D}{Dt}\left(\mu^{a}\frac{\partial F_{a}}{\partial\ddot{q}^{i}}\right),
\]
where the symbol $\frac{D}{Dt}$ denotes the total derivative with
respect to the variable $t$ occurying in all arguments of functions
$F_{a}.$ The point is that the second derivatives of of $q^{i}$with
respect to $t$ are very important here. But in the ``usual'', i.e.,
constraints-free theory those acceleration variables decideabout the
structure of equations of motion. Therefore, for $N\geq2$ the dependence
on acceleration variables becomes much more complicated. Some kind
of exception occurs only when $F_{a-s}$ are linear in second derivatives
$\ddot{q}^{i}$. Then again $R_{i}$-s do not modify essentially the
structure of second order differential equations for $q^{i}(t)$.

\section{Models of rotation-less motion}

There is some interesting question concerning non-holonomic constraints
imposed on the affine motion\cite{Gol_03}. By this we mean the strange
and interesting problem of rotation-less motion. We describe this
motion in the way that the affine velocity $\Omega^{i}{}_{j}$ is
$g$-symmetric. 
\begin{equation}
\Omega^{i}{}_{j}-\Omega_{j}{}^{i}=\Omega^{i}{}_{j}-g_{jk}g^{il}\Omega^{k}{}_{l}=0.\label{EQ95}
\end{equation}
It is well known that symmetric matrices do not form a Lie algebra.
Moreover they induce non-holonomic constraints and they are not integrable
to any sub-manifold.

Let us remind that the affine motion is defined by: 
\begin{equation}
\xi^{i}(t)=r^{i}(t)+\varphi^{i}{}_{A}(t)a^{A},
\end{equation}
where $\xi^{i}$ are Euler coordinates and $a^{A}$ are Lagrange coordinates
\cite{Burov_96}--\cite{Wulf-Rob_02}. The affine velocity is given
as follows: 
\begin{equation}
\Omega^{i}{}_{j}:=\frac{d\varphi^{i}{}_{A}}{dt}\left(\varphi^{-1}\right)^{A}{}_{j},\label{EQ96}
\end{equation}
and its co-moving description: 
\begin{equation}
\widehat{\Omega}^{A}{}_{B}=\left(\varphi^{-1}\right)^{A}{}_{i}\frac{d\varphi^{i}\!_{B}}{dt}=\left(\varphi^{-1}\right)^{A}{}_{i}\Omega^{i}{}_{j}\varphi^{j}{}_{B}.\label{EQ97}
\end{equation}

We use the polar decomposition of $\varphi$: 
\begin{equation}
\varphi=UA,\qquad\eta_{AB}=g_{ij}U^{i}\!_{A}U^{j}\!_{B},\qquad\eta_{AC}A^{C}\!_{B}=\eta_{BC}A^{C}\!_{A},\label{EQ98}
\end{equation}
where $U$ are orthogonal (isometric) matrices, i.e., $U\in O\left(U,\eta;V,g\right)$,
and $A$ are $\eta$-symmetric, i.e., $A\in\mathrm{Symm}\left(U,\eta\right)$,
and positively definite. The co-moving angular velocity $\widehat{\omega}$
of the $U$-rotator is as follows: 
\begin{equation}
\widehat{\omega}=U^{-1}\frac{dU}{dt},\label{EQ99}
\end{equation}
of course $\widehat{\omega}$ is $\eta$-skew-symmetric: 
\begin{equation}
\eta_{AC}\widehat{\omega}^{C}\!_{B}=-\eta_{BC}\widehat{\omega}^{C}\!_{A}.\label{EQ100}
\end{equation}
The kinetic energy is given in the form: 
\begin{equation}
T=T_{tr}+T_{int}=\frac{m}{2}g_{ij}\frac{dr^{i}}{dt}\frac{dr^{j}}{dt}+\frac{1}{2}g_{ij}\frac{d\varphi^{i}{}_{A}}{dt}\frac{d\varphi^{j}{}_{B}}{dt}J^{AB},\label{EQ101}
\end{equation}
where $m$ and $J^{AB}$ are the total mass and co-moving tensor of
inertia respectively given as follows: 
\begin{equation}
m=\int d\mu(a),\qquad J^{AB}=\int a^{A}a^{B}d\mu(a).
\end{equation}

In polar decomposition (\ref{EQ98}) the internal kinetic energy $T_{int}$
becomes 
\begin{eqnarray}
T_{int} & = & \frac{1}{2}\eta_{KL}\frac{dA^{K}\!_{A}}{dt}\frac{dA^{L}\!_{B}}{dt}J^{AB}+\eta_{KL}\widehat{\omega}^{K}\!_{C}A^{C}\!_{A}\frac{dA^{L}\!_{B}}{dt}J^{AB}\nonumber \\
 & + & \frac{1}{2}\eta_{KL}\widehat{\omega}^{K}\!_{C}\widehat{\omega}^{L}\!_{D}A^{C}\!_{A}A^{D}\!_{B}J^{AB}.\label{EQ102}
\end{eqnarray}
The symmetry constraints on $\Omega$ imply that: 
\begin{equation}
\widehat{\omega}=\frac{1}{2}\left[A^{-1},\frac{dA}{dt}\right]=\frac{1}{2}\left(A^{-1}\frac{dA}{dt}-\frac{dA}{dt}A^{-1}\right).\label{EQ103}
\end{equation}
Substituting this to $T_{int}$ we obtain the simplest vakonomic Lagrangian:
\begin{eqnarray}
L^{vak}=T_{int}^{vak}-\mathcal{V}\left(G\right) & = & \frac{1}{8}\eta_{KL}\frac{dA^{K}\!_{A}}{dt}\frac{dA^{L}\!_{B}}{dt}J^{AB}\nonumber \\
 & + & \frac{1}{4}\eta_{KL}\left.A^{-1}\right.^{K}\!_{D}\frac{dA^{D}\!_{C}}{dt}A^{C}\!_{A}\frac{dA^{L}\!_{B}}{dt}J^{AB}\nonumber \\
 & + & \frac{1}{8}\!\eta_{KL}\!\left.A^{\!-1}\right.^{K}\!\!\!_{E}\!\frac{dA^{\! E}\!\!_{C}}{dt}A^{\! C}\!\!\!_{A}\left.A^{\!-1}\right.^{L}\!_{F}\!\frac{dA^{F}\!\!\!_{D}}{dt}A^{\! D}\!\!\!_{B}J^{AB}-\mathcal{V}\left(G\right),\qquad\label{EQ104}
\end{eqnarray}
where $\mathcal{V}(G)$ is the potential depending only on the Green
deformation tensor: 
\begin{equation}
G_{AB}=g_{ij}\varphi^{i}\!_{A}\varphi^{j}\!_{B}=\eta_{CD}A^{C}\!_{A}A^{D}\!_{B}.\label{EQ108}
\end{equation}
The above Lagrangian is expressed through $(A,dA/dt)$. We can subject
it to the variational procedure by the direct substitution: $A\rightarrow A+\delta A$
and developing the resulting $\delta L$ up to first-order terms in
the $\eta$-symmetric $\delta A$.

The structures of vakonomic and d'Alembert equations are different.
Details will be disscussed in a forthcoming paper. Obviously, the
general d'Alembert Lagrangian has the form: 
\begin{equation}
L=T_{int}-\mathcal{V}(G)\label{EQ109}
\end{equation}
where $T_{int}$ is given by (\ref{EQ102}) and the constraints (\ref{EQ103})
are eliminated through the d'Alembert procedure of the ideal reactions.
Roughly speaking, Lagrangian equations of motion for the ``d'Alembert-constrained''
system are more complicated and given by the symmetric part of equation
of motion without constraints {[}13{]}. What is incomparatively more
essential, the d'Alembert equations of motion are obtained in different
way: firstly one takes the symmetric part of the tensorial equations
of motion and only then one substitutes to them the condition $\Omega^{i}{}_{j}=\Omega_{j}{}^{i}$.
In vakonomic theory the sequence of procedure is reversed: first one
substitutes the symmetry condition to the Lagrangian, only then one
performs the variation with respect to the symmetric $\Omega^{ij}$.
The two procedures are non-commutative. This is evidently something
else than the procedure obtained from (\ref{EQ104}) by the method
described after it, just the special case of the difference between
(\ref{EQ36}) and (\ref{EQ52})

The vakonomic model from mathematical point of view seems to be more
elegant, but it is hard to say that it is better in usual non-holonomic
problems of the slide-free motion. However, it is still being a promising
procedure, which is applicable in many areas (e.g., finance, robotics,
control, biological and nuclear problems). In general they seem to
be useful for systems with higher disorder.

\section*{Acknowledgements}

This paper partially contains results obtained within the framework
of the research project N N501 049 540 financed from the Scientific
Research Support Fund in the years 2011-2014. The authors are greatly
indebted to the Polish Ministry of Science and Higher Education for
this financial support. I'm greatful to Proffesor J. J. S\l{}awianowski
for consultations concerning constreints, and couraging me to get
into this topic.

\end{document}